\RequirePackage{etex}
\documentclass[a4paper,twocolumn,10pt,unpublished,floatfix]{quantumarticle}
\pdfoutput=1

\usepackage{env}

\title{A Photonic Parameter-shift Rule: Enabling Gradient Computation for Photonic Quantum Computers
}
\author[1]{Axel Pappalardo}
\author[1]{Pierre-Emmanuel Emeriau}
\email{pe.emeriau@quandela.com}
\author[2]{Giovanni de Felice}
\author[1]{Brian Ventura}
\author{Hugo Jaunin}
\author[2]{Richie Yeung}
\author[2]{Bob Coecke}
\author[1]{Shane Mansfield}
\affil[1]{Quandela SAS, 7 Rue Léonard de Vinci, 91300 Massy, France}
\affil[2]{Quantinuum, 17 Beaumont street, Oxford, OX1 2NA, United Kingdom}

\begin{document}

\maketitle

\begin{abstract}
We present a method for gradient computation in quantum algorithms implemented on linear optical quantum computing platforms.
While parameter-shift rules have become a staple in qubit gate-based quantum computing for calculating gradients, their direct application to photonic platforms has been hindered by the infinite dimensionality of phase-shift operators in Fock space. 
We introduce a photonic parameter-shift rule that overcomes this limitation, providing an exact formula for gradient computation in linear optical quantum processors.
Our method scales linearly with the number of input photons and utilizes the same parameterized photonic circuit with shifted parameters for each evaluation. This advancement bridges a crucial gap in photonic quantum computing, enabling efficient gradient-based optimization for variational quantum algorithms on near-term photonic quantum processors. We demonstrate the efficacy of our approach through numerical simulations in quantum chemistry and generative modeling tasks, showing superior optimization performance as well as robustness to noise from finite sampling and photon distinguishability compared to other gradient-based and gradient-free methods.

\end{abstract}

\section{Introduction}

Quantum algorithms can offer significant speedups and novel capabilities beyond their classical counterparts \cite{montanaro2016quantum}.
Variational quantum algorithms (VQAs) \cite{cerezo_variational_2021} have emerged as one of the major families of quantum algorithms and have garnered considerable interest in recent years.
While many well-known quantum algorithms are expected to require fault-tolerant quantum computers to exhibit advantages over their best classical counterparts, VQAs are considered to be more promising for more near-term Noisy Intermediate-Scale Quantum (NISQ) technologies due to the inherent robustness of variational algorithms. At the same time, their interest extends into the large-scale fault-tolerant regime for quantum computing too, whether in their own right or in synergy with more demanding algorithms. An example is the Variational Quantum Eigensolver \cite{Peruzzo_2014} which first emerged as a possible technique for estimating initial states for the Quantum Phase Estimation algorithm \cite{kitaev1995quantum}, thus combining the resilience of variational methods with the precision of more demanding fault-tolerant algorithms.

Photonic technologies are among the most promising platforms being pursued for quantum computing~\cite{rudolph2017optimistic,pelucchi2022potential,arrazola2021}. While their light footprint due to lower cryogenic requirements compared to other technologies has long lent itself to first demonstrations and proofs-of-concept in the field~\cite{bouwmeester_experimental_1997,o2003demonstration,walther2005experimental,Ascella,bao2023}, these same advantages together with native networkability have also been leveraged to deliver detailed blueprints for large-scale fault-tolerant quantum computing \cite{li2015, herr2018, auger2018, bartolucci2023,PhysRevLett.131.120603,deGliniasty2024spinopticalquantum}. Photonic models have been crucial in the conceptualization \cite{aaronson2010computational} and early demonstrations of quantum computational advantages \cite{pan_advantage_2020, madsen_quantum_2022}. Recent advances have also delivered powerful software tools \cite{heurtel2023perceval} and cloud-accessible reprogrammable photonic quantum processors \cite{Ascella}. The latter provide an ideal testing ground for photonic quantum algorithms, in particular photonic VQAs.


VQAs share a common architecture: a parameterized quantum circuit (PQC) is executed to measure expectation values or probability distributions, which are then fed into a classical device to compute a cost function. 
The classical device employs an optimisation scheme to adjust the PQC parameters, aiming to minimize this cost function iteratively.
Due to the inherent noise in NISQ devices, approximate
gradient-based methods like finite differences are often unsuitable for VQAs as they are highly sensitive to noise and prone to inaccurate gradient estimation  \cite{ORBi-ddaa8318-f2a8-4a46-9908-f2c6781a7199}. 
As a result, gradient-free optimisation methods, such as COBYLA and Nelder-Mead \cite{Cobyla}, are typically preferred for their robustness against noisy evaluations. 
However, gradient-free optimisers, while more robust against noise, often suffer from slow convergence, poor scalability, and a tendency to get stuck in local minima due to their lack of gradient information, making them inefficient for high-dimensional or complex optimisation landscapes.

For gate-based quantum circuits, a popular solution is the use of so-called parameter-shift rules (PSRs) \cite{mitarai2018quantum,schuld2019evaluating,crooks2019gradients,Wierichs2022generalparameter}. PSRs are methods that allow the exact computation of gradients of a loss function computed from a parameterized quantum circuit by evaluating multiple instances of the same circuit with shifted parameters and combining these evaluations in a structured way. 


However, these techniques do not directly extend to the linear optical setting. Photonic quantum circuits are typically parameterized by tunable phase-shifters, represented in the Heisenberg picture by the map $M \mapsto e^{-i\hat{n}\theta} M e^{i\hat{n}\theta}$ where $\theta$ is the parameterized angle,  $\hat{n}$ is the number operator and $M$ is an observable on the Fock space.  Differentiating this operator with respect to $\theta$ yields $M \mapsto e^{-i\hat{n}\theta} i[M, \hat{n}] e^{i\hat{n}\theta}$, which is not unitary and more crucially unbounded. This prevents it from being a physically realizable operation within linear optics on the same number of modes.

Here we bridge the gap by formalising a photonic parameter-shift rule, which provides an exact formula for computing the gradient of the expectation values of linear optical circuits in arbitrary states with a finite total number of photons.
The method requires a number of circuit evaluations depending linearly on the number of input photons, each using the same parameterized photonic circuit with shifted parameters. 
We show analytically that the photonic PSR is robust against finite sampling noise, quantifying the advantage compared to finite difference methods: using the PSR can provide orders of magnitude improvement in the number of samples required to reach a given precision for estimating the gradient.
We numerically validate our results by comparing its performance against widely used optimisation methods on two VQA tasks in the presence of partial distinguishability and finite sampling: a variational quantum eigensolver (VQE) and a quantum circuit born machine (QCBM). The latter features the adaptation of the shift rule to more complex loss functions appearing in machine learning, Kullback-Leibler divergence and Maximum Mean Discrepancy, showing the wide applicability of our method. 

\textbf{Prior and related works.}
In \cite{defelice2024differentiation}, the authors employ unitary dilation to address non-unitarity coming from differentiating the circuit into another linear optical circuit. 
However, this approach doubles the number of modes in the circuit and requires an additional single photon, deviating from the principle of parameter-shift rules, which rely on reusing the same circuit with shifted parameters. Moreover, the dilation method induces exponential costs in finite sampling, as amplitudes need to be rescaled. 
These constraints are particularly significant in the NISQ regime, where devices have limited size and capabilities.


In \cite{cimini2024variational}, the authors present a similar parameter shift rule for linear optical circuits,
that they use for experimental quantum metrology.
The same shift-rule appears in very recent work of Facelli et al. \cite{facelli2024exact}, 
whose results have more overlapping with ours.
Their derivations both follow Wierich et al. \cite{Wierichs2022generalparameter}
which gives a trigonometric expression of the shift rule through Dirichlet kernels.
Our derivation is more direct and does not make any assumption about the observable. 
We express the derivative with the commutator as in \cite{schuld2019evaluating} and 
find a more general characterisation, showing that photonic shift rules correspond to solutions of Equation~\eqref{eq:equation for the coefficients}.

Other distinguishing aspects of our work include practical Hoeffding bounds for the number of samples in gradient estimation that do not require access to the moment generating function, extension of the photonic PSR to more complex loss functions, and the empirical comparison to gradient-free and finite difference methods, showing the robustness of the photonic PSR to both shot noise and photon distinguishability. 

\section{Computing exact gradients of linear optical circuits}
\label{sec:PSR}

In the context of VQAs, we consider cost functions built from expectation values of the form:
\begin{equation}\label{expectation-value}
    f(\theta) = \braket{\psi|U^{\dagger}(\theta) M U(\theta)|\psi}
\end{equation}
where $\ket{\psi}$ is the quantum state on which the parameterized unitary $U$ is applied with parameter $\theta$ and where the observable $M$ is measured. 
Informally, the aim of PSRs is the exact computation of $\partial_\theta f$ from evaluation of $f$ at shifted values.

\subsection{Derivation of a photonic PSR}
\label{subsec:photonicPSR}

We now focus on expectation values coming from linear optical circuits. 
Let us set the optical circuit (a typical example is shown in Fig.~\ref{fig:causal cone}) to be composed of $m$ optical modes with $n$ single photons at the input. 
That is, we work in the Fock space for $n$ photons in $m$ modes. 
The input state is assumed to be any state with total number of photons $n$, that is $\ket{\psi} \in \text{span}(\Phi_n^m) = \text{span}(\{ \ket{s_1,...,s_m} \quad \text{with} \quad \sum_{i=1}^m s_i = n \})$.

We consider \textit{a single parameter} $\theta$ corresponding to the angle of a phase shifter in mode $k$.\footnote{The generalisation to multiple parameters is easily obtained by the chain rule.}
In this case, the unitary representing the circuit can be written as 
$W_1 e^{i\hat n_k \theta} W_2$ where $W_{1,2}$ are linear optical unitaries independent of $\theta$ and $\hat n_k$ is the number operator on mode $k$.
Then we have that:
\begin{equation}
\begin{split}
    f(\theta) & = \braket{\psi| W_2^\dagger e^{-i\hat n_k \theta} W_1^\dagger M W_1 e^{i\hat n_k \theta} W_2 |\psi}\\ 
    & = \braket{\psi' |e^{-i\hat n_k \theta} {M'} e^{i\hat n_k \theta}|\psi'}
\end{split}
\end{equation}
where $\ket{\psi'} = W_2 \ket{\psi}$ and $ {M'} = W_1^\dagger M W_1$. We now drop the apostrophe for simplicity.
By the product rule, the derivative of $f$ with respect to $\theta$ can be expressed as:
\begin{equation}
    \partial_\theta f(\theta) = i \braket{ \psi|e^{-i\hat n_k \theta} [M ,\hat n_k]  e^{i\hat n_k \theta}| \psi}
\end{equation}

To obtain a valid PSR \cite{schuld2019evaluating}, we are looking for scalars $(c_p)_{p \in \llbracket0,P\rrbracket} \in \mathbb{C}^P$ and angles $(\theta_p)_{p \in [0,P]} \in [0,2\pi)^P$ for some integer $P \in \mathbb{N}^*$ such that:

\begin{equation} \label{eq:condition on cp}
    i [M,\hat n_k] = \sum_{p=1}^P c_p e^{-i\hat n_k \theta_p} M e^{i\hat n_k \theta_p}
\end{equation}






By ordering the Fock basis $\Phi_n^m$ such that we have first all states with 0 photons in mode $k$, then all states with 1 photon in mode $k$, etc., up to the states with $n$ photons in mode $k$, the action of $e^{i\hat n_k \theta}$ can be represented on the Fock space by the following matrix:

\[
\text{diag}(\underbrace{1, \dots ,1}_\text{0 photon in mode k},\underbrace{e^{i\theta}, \dots ,e^{i\theta}}_\text{1 photon},\underset{...}{.....},\underbrace{e^{in\theta}, \dots, e^{in\theta}}_\text{n photons}) .
\]

\noindent
We also express $M$ in the Fock basis with the same ordering and we write its coefficients $(m_{ij})_{i,j \in \llbracket0,C_n^{m+n-1}\rrbracket}$ where $C_k^n$ are the binomial coefficients.

We proceed by element-wise identification. We start by computing the matrix elements $([M,\hat n_k])_{lj}$ and $ (e^{-i\hat n_k \nu} M e^{i\hat n_k \nu})_{lj}$:

\begin{equation}
([M,\hat n_k])_{lj} = ((n_k)_{jj}-(n_k)_{ll})m_{lj}
\end{equation}

\begin{equation}
(e^{-i\hat n_k \nu} M e^{i\hat n_k \nu})_{lj} = e^{i\nu ((n_k)_{jj}-(n_k)_{ll})}m_{lj} .
\end{equation}

\noindent
and then we identify from Eq.~\ref{eq:condition on cp}: 

\begin{equation}\sum_{p=1}^P c_p e^{i\theta_p ((n_k)_{jj}-(n_k)_{ll})}m_{lj} = i((n_k)_{jj}-(n_k)_{ll})m_{lj} .\end{equation}

\noindent
We can express $(n_k)_{jj}$ as a sequence $(a_j)_{j\in \mathbb{N}}$, where $a_j \in \llbracket0,n\rrbracket$ is the number of photons in mode $k$ of the $j^{th}$ element of $\Phi_n^m$.
Since $a_j \in \llbracket0,n\rrbracket$, we have $(a_{j}-a_{l}) \in \llbracket-n,n\rrbracket$,

\begin{equation}
\sum_{p=1}^P c_p e^{i\theta_p (a_j-a_l)}m_{lj} = i(a_j-a_l)m_{lj} . 
\end{equation}

\noindent
Given that the previous result holds for any observable $M$, we have:

\begin{equation} \label{eq:equation for the coefficients}
    \sum_{p=1}^P c_p e^{ij\theta_p} = ij \; ,\quad \forall j\in \llbracket-n,n\rrbracket .
\end{equation}

The system Eq.~\eqref{eq:equation for the coefficients} is overparameterized and we can thus derive multiple PSRs.
One possible solution can be obtained by fixing the $\theta_p$ and solving the resulting linear system for a sufficiently large $P \in \mathbb{N}^*$. 
A canonical choice is to fix $P= 2n$ and the angles to be $\theta_p = \frac{2\pi p}{2n+1}$. Let $\omega = e^{\frac{2i\pi}{2n+1}}$ -- the $(2n+1) ^\text{th}$ roots of unity.  Eq.~\eqref{eq:equation for the coefficients} can be rewritten as the following linear system:
\begin{equation*}
\scalebox{.95}{$
\underbrace{
    \begin{bmatrix}
    1 & 1 & 1 & \cdots & 1 \\
    1 & \omega & \omega^2 & \cdots & \omega^{2n} \\
    \vdots & \vdots & \vdots & & \vdots \\
    \vdots & \vdots & \vdots & & \vdots \\
    1 & \omega^{2n} & \omega^{2(2n)} & \cdots & \omega^{2n(2n)}
    \end{bmatrix}
    }_\text{DFT Matrix}
    \times
    \begin{bmatrix}
        c_0 \\
        c_1 \\
        \vdots \\
        c_n \\
        c_{n+1} \\
        \vdots \\
        c_{2n}
    \end{bmatrix}
    =
    i
    \begin{bmatrix}
        0 \\
        1 \\
        \vdots \\
        n \\
        -n \\
        \vdots \\
        -1
    \end{bmatrix} .$}
\end{equation*}
\noindent
Solving the linear system for $(c_p)_p$, we get that they are the inverse Discrete Fourier Transform (DFT) of $i(0,1,...,n,-n,...,-1)^T$ with $c_0 = 0$.
We thus obtain the following photonic PSR: 
\begin{equation}
\label{eq:shift_rule}
\begin{split}
    \partial_\theta f(\theta) & = \sum_{p=1}^{2n} c_p \braket{\psi|e^{-i\hat n_k (\theta + \theta_p)} M e^{i\hat n_k (\theta + \theta_p)}|\psi} \\
    & = \sum_{p=1}^{2n} c_p f(\theta + \theta_p)\; .
\end{split}
\end{equation}

Remark that Eq.~\eqref{eq:shift_rule} is an exact equation and not an approximation: through macroscopic shifts, the photonic PSR is able to capture the exact gradient. 
Our derivation requires 2n evaluations per gradient computation. 
This straightforwardly extends to multiple parameter by combining individual photonic PSR.

This is one possible formulation of the photonic PSR with a choice of $P$ and $\theta_p$'s but one could derive other photonic PSRs as long as Eq.~\eqref{eq:equation for the coefficients} is satisfied. 
It is an open question to know whether the solution based on the Fourier transform is optimal in the number of evaluations and the application one considers. 

\subsection{Robustness to finite-sampling}
\label{subsec:finitesampling}

We denote by $N_{\text{PSR}}$ the number of samples to estimate the gradient via the PSR with additive error $\epsilon$ and by $N_{\text{FD}}$ the number of samples to estimate the gradient via the finite difference with an additive error $\epsilon + \Delta$ where $\Delta$ is the chosen stepsize. 

We are interested in the number of samples needed to reach a given precision on the gradient for both the photonic PSR and finite differences. 
By applying Hoeffding's inequality to the two gradient estimators derived from the photonic PSR and finite difference methods,
the number of samples required to reach the above precision for estimating the gradient obeys:
\begin{equation}
    \frac{N_{\text{FD}}}{N_{\text{PSR}}} = \frac{4 }{(\sum_{p=1}^P |c_p|)^2 \Delta^2 } \; .
\end{equation}
A full proof is provided in Appendix~\ref{app:comparisonsample}.

Interestingly, this ratio depends only on $\Delta$ -- the stepsize for finite difference -- and $(\sum_{p=1}^{P} |c_p|)^2$ which is only determined by the number of photons in the input state. We observe numerically that this quantity scales as $n^{\alpha}$ with $\alpha \le 2.3$.

To put this result into context, considering only sampling noise and assuming the error from the Taylor approximation is negligible with $\Delta = 0.01$ (this is beneficial to finite-differences), achieving an additive error on the true gradient of $0.1$ with at least 90\% confidence for a 4-photon circuit would require approximately $26.5 \times 10^3$ samples to compute the gradient of one parameter using the photonic PSR. 
In contrast, using finite differences would necessitate around $24.0\times 10^6$ samples -- an improvement of three orders of magnitude. 


\subsection{Practical reduction of the number of evaluations}
\label{subsec:lightcone}

\begin{figure}[h]
\includegraphics[scale=0.35]{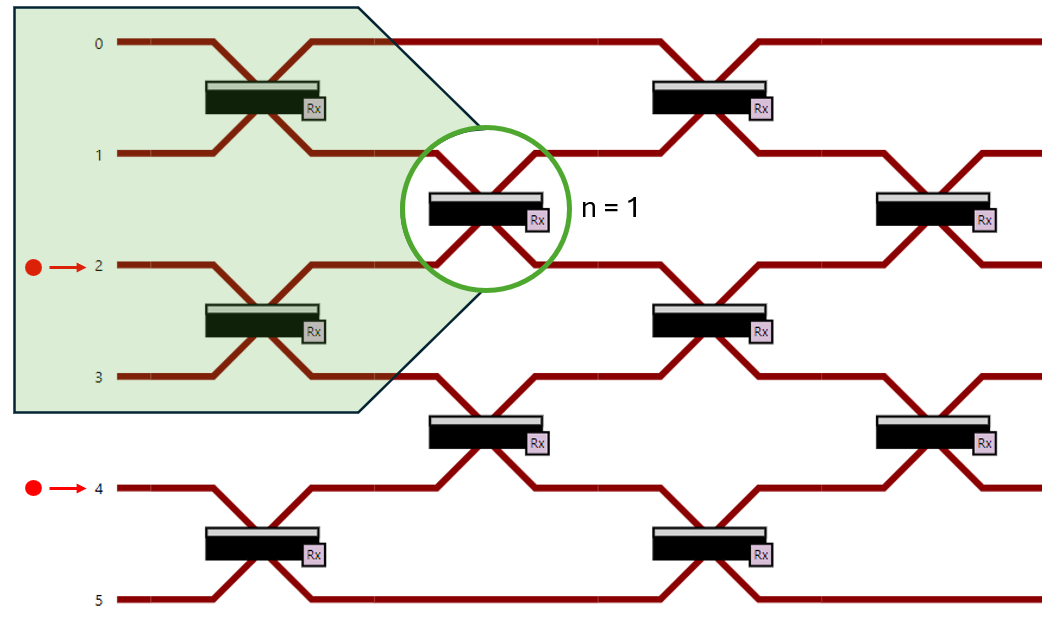}
\centering
\caption{Example of a causal cone for a circuit with 6 modes and 2 single photons at input. In this example, to compute the gradient of the highlighted parameterized gate with the PSR, one would only need to evaluate 2 terms out of 4 terms. This is because only one photon is in the past light cone of the phase-shifter.}
\label{fig:causal cone}
\end{figure}

An important practical reduction of the number of evaluations of the photonic PSR can be achieved when considering the maximum number of photons that could travel through the tunable phase-shifter one wants to differentiate. Indeed, by \textit{a light cone argument} (see Fig.~\ref{fig:causal cone}), it is possible to bound the maximum number of photons travelling through a given phase-shifter and thus reduce the number of evaluations in accordance with Eq.~\eqref{eq:shift_rule}.

\section{Applications to VQAs}
\label{sec:applications}

We use the software tool \textit{Perceval} \cite{heurtel2023perceval} to make all numerical simulations. 

\subsection{VQE}
\label{subsec:VQE}

\begin{figure*}[t]
\centering
    \subfloat[$N = + \infty$, HOM $= 1$]{\includegraphics[width = 3in]{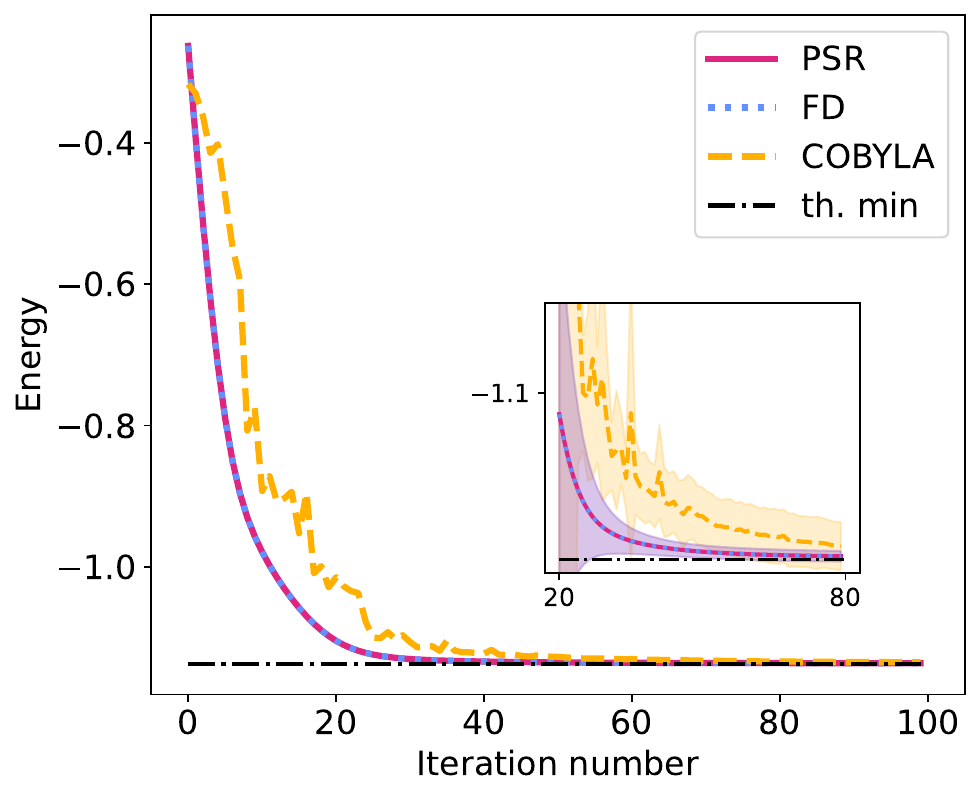}
    \label{subfig: VQE a}} 
    \subfloat[$N = 5000$, HOM $= 1$]{\includegraphics[width = 3in]{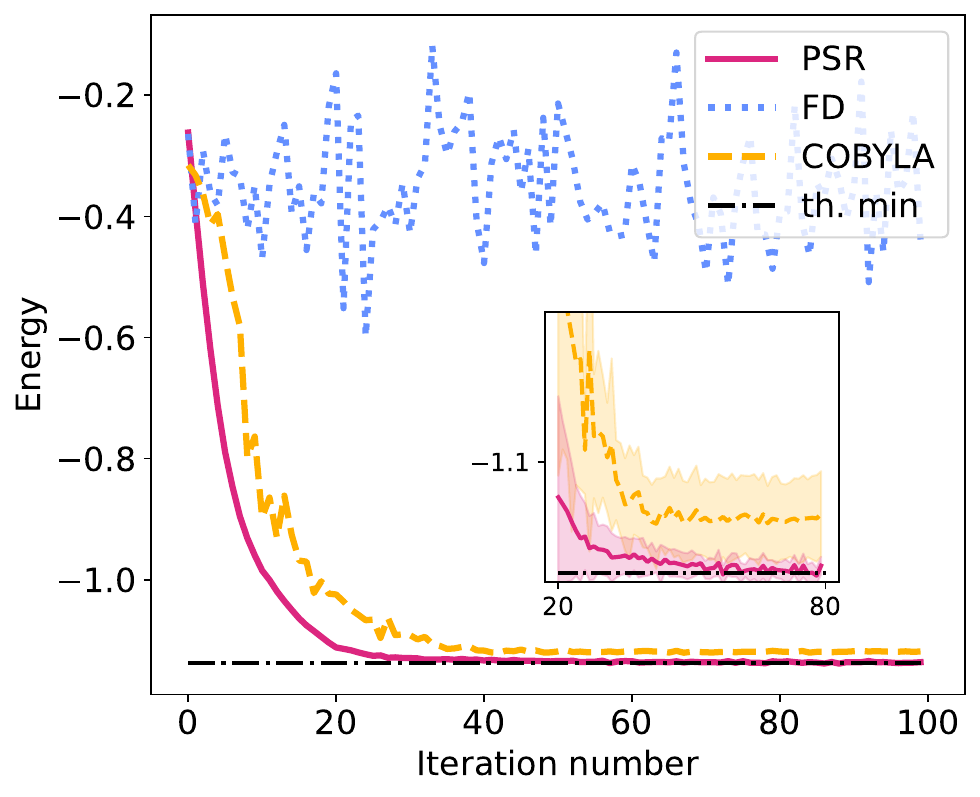}\label{subfig: VQE b}} \\
    \subfloat[$N = + \infty$, HOM $= 0.9$]{\includegraphics[width = 3in]{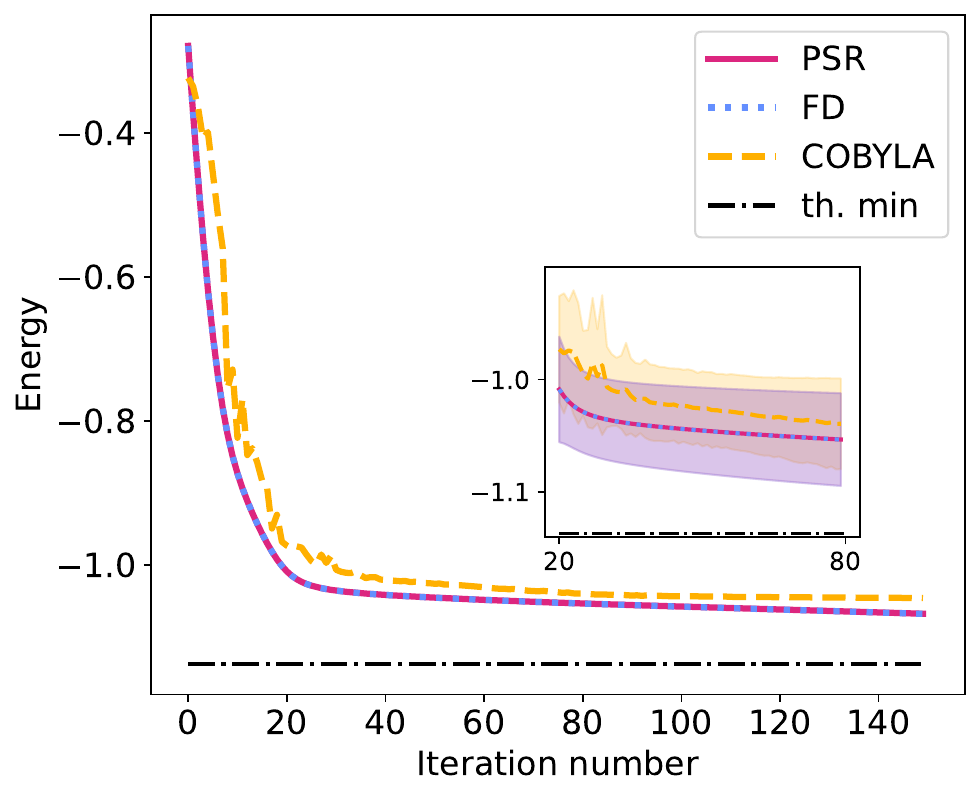}\label{subfig: VQE c}}
    \subfloat[$N = 5000$, HOM $= 0.9$]{\includegraphics[width = 3in]{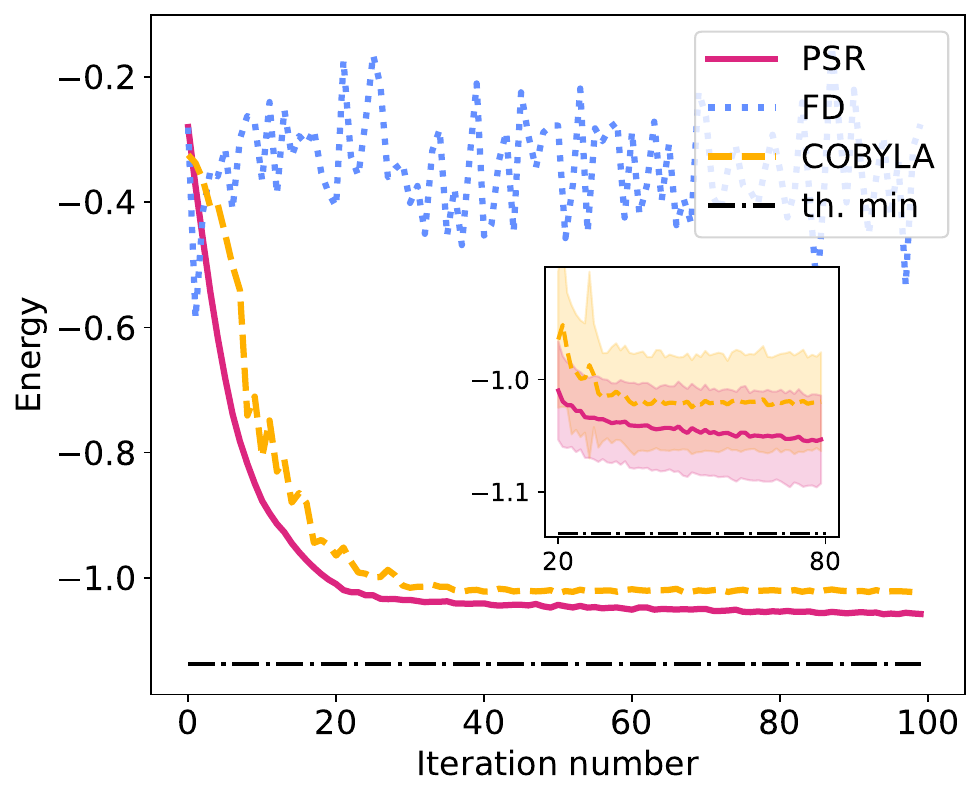}\label{subfig: VQE d}} 
\caption{Evolution of the loss function on a simple example of VQE to estimate the ground state energy of the H$_2$ molecule with three different optimisation methods: gradient descents based on the photonic PSR (solid red), finite-differences  (dotted blue) and the gradient-free method COBYLA (dashed yellow). 
The results are averaged over 10 different initial conditions, with the shaded regions in the inset plots representing the standard deviation.
(a) No finite sampling, perfect indistinguishability. (b) 5000 samples, perfect indistinguishability. (c) No finite sampling, indistinguishability of 90\%. (d) 5000 samples, indistinguishability of 90\%.
}
\label{fig:VQE results}
\end{figure*}

We first compare gradient computations for a Variational Quantum Eigensolver (VQE) algorithm to compute the ground-state energies of an H$_2$ molecule \cite{PhysRevX.8.011021}. 
A PQC produces an ansatz which is used to evaluate the energy of a given Hamiltonian, which is encoded in accordance with the ansatz. 
A classical optimisation method then iteratively updates the parameters of the PQC to converge to the ground state energy of the Hamiltonian. 

We choose this framework to showcase how gradient descent behaves when using a photonic parameter-shift rule for a well-known problem, where we can compare different optimization schemes.
We use the 2-qubit circuit used in \cite{Peruzzo_2014} for the quantum state preparation and the measurements with 8 tunable parameters. Note that, in this setting, the qubit-based parameter-shift rule \cite{schuld2019evaluating,mitarai2018quantum} works so it provides a good test case.

We denote by $N$ the number of samples used to estimate one expectation value and HOM the indistinguishability (or Hong-Ou-Mandel visibility) of the photons. The HOM is a quantity between 0 and 1 where 1 indicates that the photons are perfectly indistinguishable (the ideal, noise-free case) and 0 when they are totally distinguishable (maximal noise where the photons behave classically). When we want to see the effect of finite sampling, we set $N$ to 5000 as this is a realistic value for a NISQ type experiment. For partial distinguishability, we set HOM $= 0.9$ since this is a standard value for single-photon emitted from quantum dots \cite{Ascella}.

The optimisers that we compare are:
\begin{itemize}
    \item gradient descent using finite difference with stepsize $\Delta = 0.01$ and a learning rate of 0.4;
    \item gradient descent using a photonic PSR with a learning rate of 0.4;
    \item COBYLA \cite{Cobyla}, a gradient-free method with standard hyperparameters to control the trust region since they were producing the best results.\footnote{Other gradient-free methods such as Nelder-Mead were also tested, however they were excluded from Figure \ref{fig:VQE results} for clarity, since they didn't provide any substantially different results.}
\end{itemize}

The results are summarized in Figure \ref{fig:VQE results}.
We observe that, in the absence of finite sampling, finite differences and the PSR yield very similar results. This is expected since without finite sampling the gradient computed with finite differences is close to exact when choosing a small stepsize, which leads to very similar gradient descent runs. However, we can also observe that the introduction of finite sampling drastically affects finite differences as it was not possible to converge in a reasonable number of samples, while barely impacting the PSR or COBYLA. These results were expected given the theoretical results obtained in Section \ref{subsec:finitesampling}.
Additionally, we can also observe that partial distinguishability error reduces the final precision obtained with all optimizers, with a greater effect on gradient-free optimizers such as COBYLA.

To obtain a more realistic simulation, we include both noise sources in the plot in Figure \ref{subfig: VQE d} highlighting the resilience to noise of gradient descent using a photonic PSR. Indeed, 
it combines the benefits of a gradient computation method resistant to sampling noise, with a gradient descent algorithm that is seemingly more consistent than its gradient-free counterpart when introducing partial distinguishability.

\subsection{QCBM}
\label{subsec:QCBM}

Here, we look at a photonic native problem where the usual qubit parameter-shift rule cannot be applied. 
We investigate how various optimisation schemes impact the learning in photonic Quantum Circuit Born Machines (QCBM) \cite{salavrakos2024errormitigated}. 
These generative learning models are well suited for NISQ hardware since they can be implemented on shallow circuits. 

The goal of generative learning models is to learn the underlying distribution of a dataset and then generate new samples with similar properties to the training dataset. 
For the purposes of testing the optimisation scheme with the photonic PSR, the target probability distribution is known in advance in this work, and we try to best approach the known distribution. 
Of course this is not the case in practice, where the models are intended to be used without knowing the target distribution.

Typically, QCBM use the Kullback-Leibler (KL) divergence or the Maximum Mean Discrepancy (MMD) as a loss function. 
Since they are not a linear sum of expectation values from the circuit, we need to adapt the photonic PSR.
We show in Appendix~\ref{subsec:adaptation_PSR} how to reuse the photonic PSR from Eq.~\ref{eq:shift_rule} to differentiate loss functions expressed with the KL divergence or MMD. 

Gradient descent based on finite-differences is not realistic in the finite sampling regime, see Fig.~\ref{fig:VQE results} and, indeed, they were not converging in the simulations of QCBM. 
Instead, we implement gradient descent based on Simultaneous Perturbation Stochastic Approximation (SPSA) \cite{SPSA} which is more robust in the presence of shot noise \cite{salavrakos2024errormitigated}.
The optimizers we compared here are then:
\begin{itemize}
    \item gradient descent using SPSA with a learning rate of 0.4.
    \item gradient descent using a photonic PSR with a learning rate of 0.4.
    \item COBYLA \cite{Cobyla}, a gradient-free method.
\end{itemize}

\begin{figure}[ht!]
\centering
    \subfloat{\includegraphics[width = 3in]{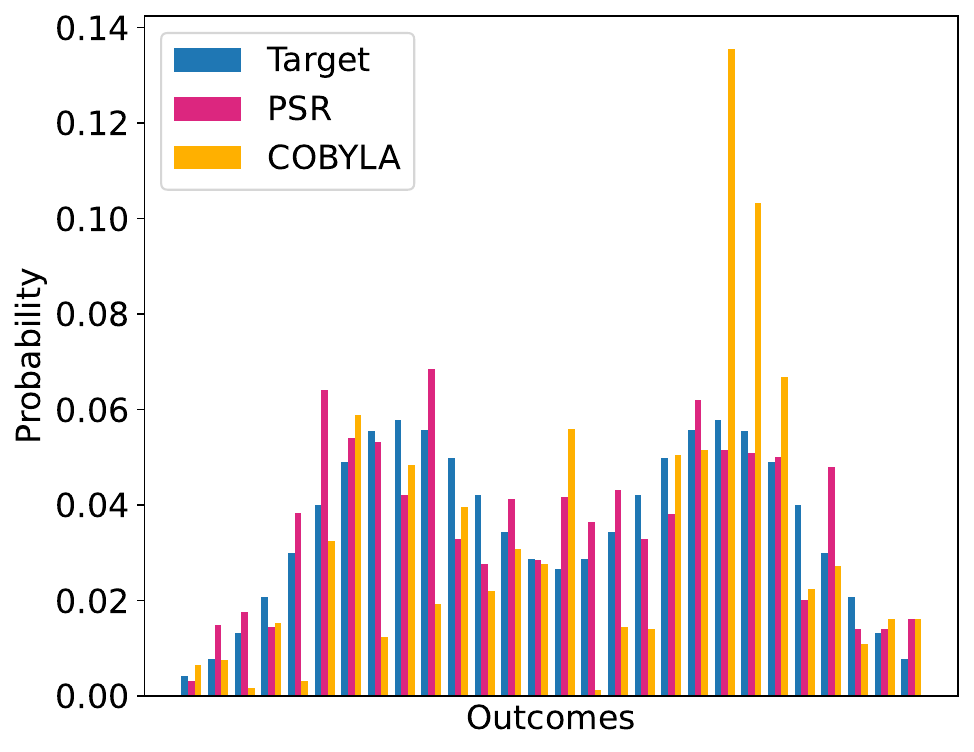}\label{subfig:pdf QCBM}}
    \\
    \subfloat{\includegraphics[width = 3in]{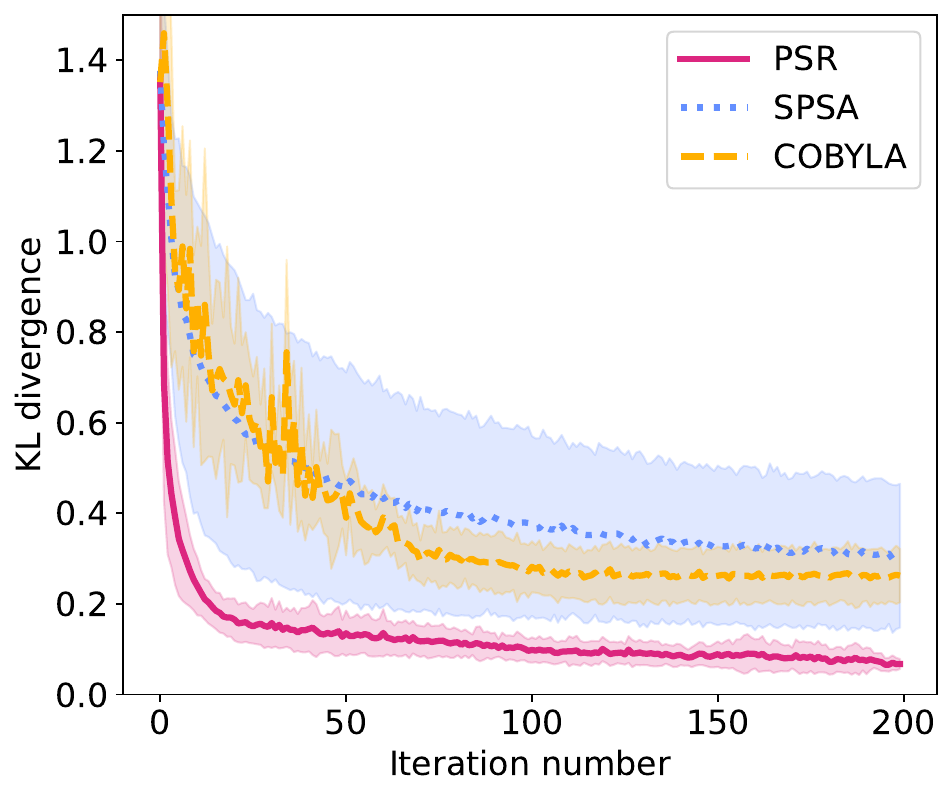}\label{subfig:QCBM b}}
\caption{QCBM numerical experiment for learning a mixture of two Gaussians with a photonic circuit comprising $n=3$ photons in $m=8$ modes with 28 tunable parameters. We have set $N=5000$ samples and HOM=0.9. (Top) Histogram obtained from the solutions based on the different optimisers and compared to the target distribution. (Bottom) Evolution of the KL divergence with different optimisation methods: PSR (solid red), SPSA  (dotted blue) and COBYLA (dashed yellow). Averages are taken over 10 runs with different initial parameters and shaded areas represented one standard deviation.}
\label{fig:QCBM results}
\end{figure}

The results are presented in Figure \ref{fig:QCBM results}.
The task is to learn a mixture of two Gaussian distributions.
Above a reasonable number of samples, PSR gradient descent is barely affected by finite sampling, while COBYLA does suffer from a low number of samples. Here we have shown a run with 5000 samples which is a typical number of samples in quantum machine learning applications.
It appears from the standard deviation observed in the different figures that SPSA is less consistent compared to the other methods, which is intuitive given its stochastic nature.
We conclude that gradient descent based on the photonic PSR provides the best precision and it converges in fewer iterations than the other methods.

\section{Discussion}
\label{sec:conclusion}

We derived a parameter-shift rule that enables gradient computation in linear optical systems.
One of the advantages of our derivation is that it does not make any assumption about the observable; i.e.\ $M$ is an arbitrary operator on the Fock space throughout this report which includes full-counting statistics \cite{ivanov2020complexity} which is not Hermitian. 
Translating our result to the Schrödinger picture, suppose that $\rho$ is a mixed state with total number of photons $n$. The phase shifter acts as $e^{i \hat n \theta} \rho e^{-i \hat n \theta}$ and its derivative with respect to $\theta$ is given by $e^{i \hat n \theta} i[\hat n , \rho] e^{-i \hat n \theta}$. From the photonic PSR, this unnormalised state can be expressed as
$\sum_{p = 1}^{2n} c_p e^{i \hat n (\theta + \theta_p)} \rho e^{-i \hat n (\theta + \theta_p)}$.
As a result, the photonic PSR can be used to compute the gradient with respect to a phase shifter of the outcome probabilities of \textit{an arbitrary quantum optical system} with a bounded number of photons.

We have analysed two simple applications in quantum chemistry and machine learning. As the scale of the problems considered grows, the optimisation task will likely face high dimensional non-convex landscapes.
This will necessitate going beyond $l_2$ geometry and finding improved metrics in optimisation methods such as quantum natural gradient descent \cite{Stokes_2020}.

Through Hoeffding bounds and numerical simulations, we demonstrated the improved performance achieved by the photonic PSR as compared to other gradient estimation methods, in the presence of both finite sampling and photon distinguishability. 
This opens the path to hardware demonstrations of VQAs with the photonic PSR in the NISQ era and beyond.

\vspace{1cm}
\textit{Acknowledgments}--- 
This work has been co-funded by the European Commission as part of the EIC accelerator program under the grant agreement 190188855 for SEPOQC project, by the Horizon-CL4 program under the grant agreement 101135288 for EPIQUE project, by the ANR-24-QUA2-007-003 for ResourcesQ project and by the  Quondensate project under the grant agreement 101130384. 

\bibliographystyle{quantum}

\onecolumn\newpage
\appendix

\section{Appendices}
\label{app:main}

\subsection{Comparison of the number of samples}
\label{app:comparisonsample}

We recall that the $f(\theta) = \braket{\psi | U^{\dagger}(\theta) \hat M U(\theta) | \psi}$ and that we are interested in computing $\partial_\theta f$.
In this section, we want to obtain an upper bound on the error generated by finite sampling when computing the gradient with finite differences and with a photonic PSR.

We prove here that the number of samples required to reach the same precision for estimating the gradient obey:
\begin{equation}
    \frac{N_{\text{FD}}}{N_{\text{PSR}}} = \frac{4 }{(\sum_{p=1}^P |c_p|)^2 \Delta^2 } \; ,
\end{equation}

\noindent where we note $N_{\text{PSR}}$ the number of samples to estimate the gradient via the PSR and $N_{\text{FD}}$ the number of samples via the finite difference. $\Delta$ is the chosen stepsize for the finite difference. 

We can start by recalling Hoeffding's inequality.
Let $X_1,\hdots,X_n$ be $n$ independent random variables such that $a_i \le X_i \le b_i$ almost surely (i.e. we have that $\mathbb{P}(a_i \le X_i \le b_i) = 1$). If we then consider their sum $S_n = X_1 + \hdots + X_n$, Hoeffding's inequality bounds the probability on getting an estimate with additive error $\varepsilon > 0$:
\begin{equation}
        \mathbb{P}(|S_n - \mathbb{E}[S_n]| \ge \varepsilon) \le 2\exp{ \left( \frac{-2\varepsilon^2}{\sum_{i=1}^n(b_i-a_i)^2} \right)} \; .
\end{equation}

Fix the target additive error $\varepsilon > 0$.
Let $\{ \lambda_j \}_{j \in \llbracket1,J \rrbracket}$ be the $J$ eigenvalues of the observable $\hat M$ and $\ket{\phi_j}$ as the associated eigenstates. Since the range of the estimator plays a crucial role in Hoeffding's inequality, we note $\lambda = \underset{j}{\max} |\lambda_j|$.
We can rewrite $f$ as:
\begin{equation}
    f(\theta) = \sum_{j=1}^J\lambda_j |\braket{\phi_j | U(\theta) | \psi}|^2 .
\end{equation}

Let $N$ the total number of samples used for the estimate.
Let $(X_k(\theta))_{k\in\llbracket1,N\rrbracket}$ be $N$ independent random variables such that $X_k(\theta) = \frac{\lambda_j}{N}$ if the $k^{th}$ sample is measured in state $\ket{\phi_j}$, and following the probability distribution $\mathbb{P}(X_k(\theta) = \frac{\lambda_j}{N} ) = |\braket{\phi_j | U(\theta) | \psi}|^2 $. Also $\forall k, -\frac{\lambda}{N} \le X_k(\theta) \le \frac{\lambda}{N}$.

We define: 
\begin{equation}
    S_N(\theta) = \sum_{k=1}^N X_k(\theta) \; ,
\end{equation}
and since
\begin{equation}
\mathbb{E}[X_k(\theta)] = \frac{1}{N} \sum_{j=1}^J \lambda_j |\braket{\phi_j | U(\theta) | \psi}|^2 \; ,
\end{equation}
we have
\begin{equation}\label{eq:E=f}
\begin{split}
    \mathbb{E}[S_N(\theta)] & = \frac{1}{N}\sum_{k=1}^N\sum_{j=1}^J \lambda_j |\braket{\phi_j | U(\theta) | \psi}|^2 \\
    & = \sum_{j=1}^J \lambda_j |\braket{\phi_j | U(\theta) | \psi}|^2 \\
    & = f(\theta) .
    \end{split}
\end{equation}
This allows us to express $f(\theta)$ as an expectation value of a sum of independent random variables, and therefore to use Hoeffding's inequality to quantify the noise generated by finite sampling.

\textit{Finite differences.}
Using finite differences, with a stepsize $\Delta$, we define:

\begin{equation}
    Df(\theta,\Delta) := \frac{f(\theta + \Delta) - f(\theta)}{\Delta} \underset{\Delta \rightarrow 0}{=} \partial_\theta f(\theta) \; .
\end{equation}

\noindent
Let a new random variable $Y_k(\theta,\Delta) = X_k(\theta + \Delta) - X_k(\theta)$ and its sum 
$T_N(\theta,\Delta) := \sum_{k=1}^N Y_k(\theta,\Delta)$.
Since all $X_k(\theta)$ and $X_k(\theta + \Delta )$ are independent, using Equation~\eqref{eq:E=f}, we have:

\begin{equation} \label{eq:E(T)=df}
Df(\theta,\Delta)  = \frac{1}{\Delta} \mathbb{E}[T_N(\theta,\Delta)] \; .
\end{equation}

We have $-\frac{2\lambda}{N} \le Y_k(\theta,\Delta) \le \frac{2\lambda}{N}$, $\forall k$. Using Hoeffding's inequality and Equation \ref{eq:E(T)=df} we obtain the following for any $t > 0$:

\begin{equation} \label{eq:FD upper bound}
\begin{split}
    \mathbb{P}\left(|\frac{1}{\Delta}T_N(\theta,\Delta) - Df(\theta,\Delta)| \ge \varepsilon \right) & = \mathbb{P}\left(|T_N(\theta,\Delta) - \mathbb{E}[T_N(\theta,\Delta)]| \ge \Delta \varepsilon \right) \\
    & \le 2\exp \left( -\frac{2(\Delta \varepsilon)^2}{\sum_{k=1}^N (\frac{4\lambda}{N})^2}\right) \\
    & \le 2\exp \left( -\frac{\Delta^2 \varepsilon^2 N}{8\lambda^2}\right) .
\end{split}
\end{equation}

\textit{PSR.}
When using the parameter-shift rule obtained in Eq.~\eqref{eq:shift_rule} in its most general form, we have:
\begin{equation}
\partial_\theta f(\theta) = \sum_{p=1}^P c_p f(\theta + \theta_p) \; .
\end{equation}

Similarly as in the finite differences case, we define a new random variable $Z_k(\theta) = \sum_{p=1}^P c_p X_k(\theta + \theta_p)$ and their sum:

\begin{equation}
    \begin{split}
        R_N(\theta) := \sum_{k=1}^N Z_k(\theta) & = \sum_{k=1}^N \sum_{p=1}^P c_p X_k(\theta + \theta_p) \\
        & =  \sum_{p=1}^P c_p S_N(\theta + \theta_p) \; .
    \end{split}
\end{equation}
Then:
\begin{equation}
    \mathbb{E}[R_N(\theta)] = \partial_{\theta}f(\theta) \; .
\end{equation}

\noindent
Because $-\frac{\sum_{p=1}^P |c_p|\lambda}{N} \le Z_k(\theta) \le \frac{\sum_{p=1}^P |c_p|\lambda}{N}$, $\forall k$, we can apply Hoeffding's inequality:

\begin{equation} \label{eq:PSR upper bound}
    \mathbb{P}(|R_N(\theta) - \partial_{\theta}f(\theta)| \ge \varepsilon) \le 2\exp \left( \frac{-\varepsilon^2 N}{2 \lambda^2 \left(\sum_{p=1}^P|c_p|\right)^2}\right) \; .
\end{equation}

\textit{Comparison.} We can inverse both upper bounds to obtain the number of samples required to guarantee a given precision with a certain probability. We write $\Lambda$ the desired upper bound of the probability in both cases: 

\begin{equation}
\begin{split}
    N_{FD} & = -\frac{8 \lambda^2 \ln{(\Lambda/2)}}{\varepsilon^2 \Delta^2} \\
    N_{PSR} & = -\frac{2 \lambda^2 (\sum_{p=1}^P |c_p|)^2 \ln{(\Lambda/2)}}{\varepsilon^2} .
\end{split}
\end{equation}

\noindent
Therefore:
\begin{equation}
    \frac{N_{FD}}{N_{PSR}} = \frac{4 }{(\sum_{p=1}^P |c_p|)^2 \Delta^2 } \;  .
\end{equation}

\subsection{Adaptation of the photonic PSR to typical loss functions}
\label{subsec:adaptation_PSR}

\subsubsection{Kullback-Leibler divergence}

The Kullback-Leibler (KL) divergence \cite{10.1214/aoms/1177729694} is often used as a loss function in quantum machine learning applications. We recall its expression. Given two probability distributions $Q$ and $T$ defined on a sample space $\chi$:

\begin{equation}
    D_{KL}(Q||T) = \sum_{x \in \chi} Q(x) \log \left( \frac{Q(x)}{T(x)}\right) \; .
    \label{eq:KL_div}
\end{equation}
In our case, the first distribution $Q$ is given by the outputs of the QCBM circuit so $Q(x)$ is an expectation value at point $x$, while $T$ is the target distribution.
To obtain the full gradient, we can simply compute the gradient for one parameter at a time. Hence, we can fix all but one parameter $\theta$. The KL divergence then becomes:

\begin{equation}
    D_{KL}(\theta,T) = \sum_{x \in \chi} Q_\theta(x) \log \left( \frac{Q_\theta(x)}{T(x)}\right) \; .
\end{equation}

The PSR found in Equation \ref{eq:shift_rule} cannot be applied directly since it is not linear in an quantum expectation value. However, it can still be used in a very similar manner. Indeed:

\begin{equation}
    \partial_\theta D_{KL}(\theta,T) = \sum_{x \in \chi} \partial_\theta Q_\theta(x) \left(T(x) + \log \left( \frac{Q_\theta(x)}{T(x)}\right) \right) \; . 
\end{equation}
One can use the PSR to compute $\partial_\theta Q_\theta(x)$.
Typically, $Q_\theta(x)$ can be expressed as:

\begin{equation}
    Q_\theta(x) = \braket{\psi|U^\dagger(\theta) | \hat x \rangle \langle\hat x| U(\theta)|\psi} \; ,
\end{equation}
with $\ket{\psi}$ the input state, $U(\theta)$ a photonic PQC with a single tunable phase shifter and
with $\ket{\hat x}$ the state associated to the event $x$. 
We can apply the PSR to $Q_\theta(x)$:

\begin{equation} \label{eq:PSR on proba}
    \partial_\theta Q_\theta(x) = \sum_{p=1}^P c_p Q_{\theta + \theta_p}(x) \; .
\end{equation}

\noindent
Thus we can write:

\begin{equation} \label{eq:QCBM PSR}
    \partial_\theta D_{KL}(\theta,T) = \sum_{p=1}^P \sum_{x \in \chi} c_p Q_{\theta + \theta_p}(x) \left(T(x) + \log \left( \frac{Q_\theta(x)}{T(x)}\right) \right) \; .
\end{equation}

\noindent
We can note that by fixing the $\theta_p$ to the (2n+1)$^{th}$ root of the unity, both the $\theta_p$ and the $c_p$'s only depend on the number of photons.

\subsubsection{Maximum Mean Discrepancy}
\label{subsubsec:KL}

Another cost function commonly used for QCBM \cite{Liu_2018,Coyle_2020} is the Maximum Mean Discrepancy (MMD) \cite{MMD}. We recall its definition.
Given sample spaces $\chi$ and $\mathcal F$, $Q$ and $T$ two probability distributions on $\chi$ and a feature map $\phi$ such that for any random variables $x \in \chi$, we have $\phi(x) \in \mathcal{F}$, we define the MMD as follows:

\begin{equation}
\begin{split}
        \mathcal{L}_{\text{MMD}}(Q,T) & = 
        \left\lVert  \underset{x \sim Q}{\mathbb{E}}[\phi(x)] - 
         \underset{y \sim T}{\mathbb{E}}[\phi(y)] \right\rVert_\mathcal{F}^2 \\
         & = \langle \underset{x \sim Q}{\mathbb{E}}[\phi(x)] - 
        \underset{y \sim T}{\mathbb{E}}[\phi(y)],\underset{x \sim Q}{\mathbb{E}}[\phi(x)]- \underset{y \sim T}{\mathbb{E}}[\phi(y)] \rangle_\mathcal{F} \; ,
\end{split}
\end{equation}

\noindent
which is the distance between the feature means of $Q$ and $T$.
By linearity, we have:

\begin{equation}
    \langle \underset{x \sim Q}{\mathbb{E}}[\phi(x)], \underset{y \sim T}{\mathbb{E}}[\phi(y)] \rangle_\mathcal{F} = \underset{\substack{x \sim Q \\ y \sim T}}{\mathbb{E}} [\langle \phi(x),\phi(y) \rangle_\mathcal{F}] \; .
\end{equation}

Then we define a kernel function $k$ such that 
$k(x,y) = \langle \phi(x),\phi(y) \rangle_\mathcal{F}$ and we
write $\mu_Q = \underset{x \sim Q}{\mathbb{E}}[\phi(x)]$ for clarity. 
We fix $T$ the target distribution and $Q_\theta$ the output of the QCBM with a single parameter (the generalisation being obtained by linearity). 
We can develop $\mathcal{L}_{\text{MMD}}$ as:

\begin{equation}
\begin{split}
        \mathcal{L}_{\text{MMD}}(\theta) & = 
        \langle \mu_{Q_\theta}, \mu_{Q_\theta} \rangle
        - 2 \langle \mu_{Q_\theta}, \mu_T \rangle
        + \langle \mu_T, \mu_T \rangle \\
        & = \underset{x,y \sim Q_\theta}{\mathbb{E}}[k(x,y)] + \underset{x,y \sim T}{\mathbb{E}}[k(x,y)] - 2\underset{\substack{x \sim Q_\theta \\ y \sim T}}{\mathbb{E}}[k(x,y)] \\
        & = \sum_{x,y \in \chi} k(x,y) Q_\theta (x) Q_\theta(y) + \sum_{x,y \in \chi} k(x,y) T(x) T(y)  - 2\sum_{x,y \in \chi} k(x,y) Q_\theta (x) T(y) \; ,
\end{split}
\end{equation}

Once again, the photonic PSR cannot be applied directly, but similarly to \ref{subsubsec:KL} one can express the derivative $\partial_\theta \mathcal{L}_{MMD}(\theta)$ with respect to $\partial_\theta Q_\theta$ and use the photonic PSR on the latter:

\begin{equation}
        \partial_\theta \mathcal{L}_{\text{MMD}}(\theta) = \sum_{x,y \in \chi} k(x,y) [ \partial_\theta Q_\theta (x) Q_\theta(y) +  Q_\theta (x) \partial_\theta Q_\theta(y) ]
         - 2 \sum_{x,y \in \chi} k(x,y) \partial_\theta Q_\theta (x) T(y) \; . 
\end{equation}

\noindent
Since $k$ is symmetric, one can write:

\begin{equation}
    \partial_\theta \mathcal{L}_{\text{MMD}}(\theta) = 2 \sum_{x,y \in \chi} k(x,y) \partial_\theta Q_\theta (x) Q_\theta(y) - 2 \sum_{x,y \in \chi} k(x,y) \partial_\theta Q_\theta (x) T(y) \; .
\end{equation}

\noindent
Then we apply a photonic PSR to compute $\partial_\theta Q_\theta (x)$. The gradient of the MMD cost function can be expressed as:

\begin{equation}
    \begin{split}
        \partial_\theta \mathcal{L}_{\text{MMD}}(\theta) & = 2 \sum_{p=1}^P c_p 
        \left( 
        \sum_{x,y \in \chi} k(x,y)  Q_{\theta+ \theta_p}(x) Q_\theta(y) - 2 \sum_{x,y \in \chi} k(x,y)  Q_{\theta + \theta_p}(x) T(y)
        \right) \\
        & = 2 \sum_{p=1}^P c_p 
        \left(
        \underset{\substack{x \sim Q_{\theta+ \theta_p} \\ y \sim Q_\theta}}{\mathbb{E}}[k(x,y)]
        - \underset{\substack{x \sim Q_{\theta+ \theta_p} \\ y \sim T}}{\mathbb{E}}[k(x,y)]
        \right) \; .
    \end{split}
\end{equation}

\noindent
If $\mathcal{L}_{\text{MMD}}$ is parameterized by more than one parameter, the formula can easily be generalized by using the presented derivation on all parameters. 

Regarding the choice for the kernel function $k$, we decided to use the Gaussian mixture kernel in this work:
\begin{equation}
k(x,y) = \frac{1}{c}\sum_{i=1}^c \exp{-\frac{||x-y||^2}{2\sigma_i}} \; .
\end{equation}
This is a popular choice, that reveals the difference between two distributions under various scales. The MMD loss with this kernel function guarantees that it approaches zero asymptotically if and only if the model's distribution exactly matches the target distribution \cite{MMD},\cite{gretton2008kernelmethodtwosampleproblem}.

\end{document}